\documentclass[12pt]{article}
\newcommand{\dd}{\mbox{\rm d}}

\newcommand{\oo}{\over}
\newcommand{\p}{\partial}

\newcommand{\be}{\begin{equation}}
\newcommand{\ee}{\end{equation}}
\newcommand{\lbl}{\label}

\newcommand{\vs}{\vspace}
\newcommand{\hs}{\hspace}

\begin{document}
\thispagestyle{empty}

\vspace{3cm}

\begin{center} {\LARGE \bf Clifford Algebra Based Polydimensional
Relativity and Relativistic Dynamics}\footnote{Talk presented at
the IARD 2000 Conference, 26--28 June, 2000.}

\vspace{1cm}
       
Matej Pav\v si\v c

Jo\v zef Stefan Institute, Jamova 39, SI-1000 Ljubljana, Slovenia;

e-mail: Matej.Pavsic@ijs.si

\vs{3mm}

November, 2000

\vspace{2cm}

\end{center}

\vspace{.8cm}

Starting from the geometric calculus based on Clifford algebra, the idea
that physical quantities are Clifford aggregates (``polyvectors") is
explored. A generalized point particle action (``polyvector action")
is proposed. It is shown that the polyvector action, because of the
presence of a {\it scalar} (more precisely a {\it pseudoscalar}) variable,
can be reduced to the well known, unconstrained, Stueckelberg action
which involves an invariant evolution parameter. It is pointed out that,
starting from a different direction, DeWitt and Rovelli postulated the
existence of a {\it clock} variable attached to particles which serve
as a reference system for identification of spacetime points. The action
they postulated is equivalent to the polyvector action. Relativistic
dynamics (with an invariant evolution parameter) is thus shown
to be based on even stronger theoretical and conceptual foundations than
usually believed.

\newpage

\section{Introduction}

In the history
of physics it has often happened that a good new formalism contained also
good new physics waiting to be discovered and identified in suitable
experiments. Today the so called {\it Fock-Schwinger proper time formalism}
is widely recognized for its elegance and usefulness, especially
when considering quantum fields in curved spaces. There are two main 
interpretations of the formalism:

\ (i) According to the first one, it is considered merely as a useful
calculational tool, without a physical significance. Evolution in $\tau$
and the absence of the constraint is assumed to be fictitious and
unphysical. In order to make contact with physics one has to get rid
of $\tau$ in all considered expressions by integrating them over $\tau$.
By doing so one projects unphysical expressions into the physical ones,
and in particular one projects unphysical states into the physical ones.

(ii) According to the second interpretation, evolution in $\tau$ is genuine
and physical. There is indeed the dynamics in spacetime. Mass is a constant
of motion and not a fixed constant in the Lagrangian. Such an approach was
proposed by Fock $(^{1})$ and subsequently investigated by Stueckelberg
($^2$), Feynman ($^3$),
Schwinger ($^4$), Davidon ($^5$), Horwitz ($^6$), 
Fanchi ($^7$) and many others ($^{8,9}$).

In this paper I am going to show that yet another, widely investigated
formalism based on Clifford algebra brings a strong argument in favor
of the interpretation (ii). Clifford numbers can be used to represent vectors, 
multivectors and, in general, polyvectors (which are Clifford aggregates). 
They form a very useful tool for geometry. The well known equations 
of physics can be cast into elegant compact forms by using the geometric 
calculus based on Clifford algebra.

These compact forms suggest a generalization, discussed in the literature
by Pezzaglia ($^{10}$), Castro ($^{11}$) and also in ref. ($^{12}$),
that every physical quantity
is a polyvector. For instance, the momentum polyvector in 4-dimensional 
spacetime has not only a vector part, but also a scalar,
bivector, pseudovector and pseudoscalar part. Similarly for the velocity
polyvector. Now we can straightforwardly generalize the conventional
constrained action by rewriting it in terms of polyvectors. By doing so,
we obtain in the action also a term which corresponds to the scalar or
pseudoscalar part of the velocity polyvector. A consequence of such an 
extra term is that, when confining us for simplicity to polyvectors 
with the pseudoscalar and the vector part only, the variables corresponding
to  4-vector components, can all be taken as independent. After a 
straightforward procedure in which we omit the extra term in the action,
since it turns out to be just a total derivative, we obtain the Stueckelberg 
unconstrained action! This is certainly a remarkable result. The original, 
constrained action is equivalent to the unconstrained action. 
An analogous procedure can be applied also to the extended objects such as 
strings, membranes or branes in general.

After describing briefly the essence of geometric calculus based on
Clifford algebra I am going to show how relativistic dynamics (which
contains the invariant evolution parameter) emerges from the Clifford
algebra based reformulation and generalization of the theory of relativity.
Briefly I am going to touch also few other relevants subjects.

\section{Geometric calculus based on Clifford algebra}

I am going to discuss the calculus with {\it vectors} and {\it their
generalizations}\footnote{Here I shall provide a brief, simplified
introduction into the subject. A more elaborated discussion will be
provided elsewhere ($^{13}$).}.
Geometrically, a vector is an oriented line element.

How to multiply vectors? There are two possibilities:

1. {\it The inner product}
\be
      a \cdot b = b \cdot a
\lbl{1.1}
\ee
of vectors $a$ and $b$. The quantity $a \cdot b$ is a {\it scalar}.

2. {\it The outer product}
\be
      a \wedge b = - b \wedge a
\lbl{2.2}
\ee
which is an oriented element of a plane.

 The products 1 and 2 can be considered as the {\it symmetric} and the
 {\it antisymmetric} parts of the {\it Clifford product}, called also
 {\it geometric product}
 \be
        a b = a \cdot b + a \wedge b
\lbl{2.3}
\ee
where
\be
     a \cdot b \equiv {1\oo 2} (a b + b a)
\lbl{2.4}
\ee
\be
       a \wedge b \equiv {1\oo 2} (a b - b a)                        
\lbl{2.5}
\ee

This suggests a generalization to trivectors, quadrivectors, etc. It
is convenient to introduce the name $r$-{\it vector} and call $r$ its
{\it degree}: 

\setlength{\unitlength}{1cm}
\begin{picture}(12.5,4.5)

\put(1,0){\makebox(1.8,4.5)
{\begin{minipage} [t] {1.8cm}
0-vector\\
1-vector\\
2-vector\\
3-vector\\
\centerline{.\ \ }
\centerline{.\ \ }
\centerline{.\ \ }
$r$-vector
\end{minipage} }}

\put(4,0){\makebox(5,4.5)
{\begin{minipage} [t] {5cm}
\centerline{$s$} 
\centerline{$a$}
\centerline{$a \wedge b$}
\centerline{$a \wedge b \wedge c$}
\centerline{.}
\centerline{.}
\centerline{.}
\centerline{$A_r = a_1 \wedge a_2 \wedge ... \wedge a_r$}

\end{minipage} }}

\put(9.5,0){\makebox(2,4.5)
{\begin{minipage} [t] {2cm}
scalar\\
vector\\
bivector\\
trivector\\
\centerline{.\ \ \ \ }
\centerline{.\ \ \ \ }
\centerline{.\ \ \ \ }
multivector

\end{minipage} }}

\end{picture}

In a space of finite dimension this cannot continue indefinitely:
the $n$-vector is the highest $r$-vector in $V_n$ and the $(n+1)$-vector is
identically zero. An $r$-vector $A_r$ represents an oriented $r$-volume
(or $r$-direction) in $V_n$.

Multivectors $A_r$ are the elements of the {\it Clifford algebra} 
${\cal C}_n$ of
$V_n$. An element of ${\cal C}_n$ will be called a {\it Clifford number}.
Clifford numbers can be multiplied among themselves and the results are
Clifford numbers of mixed degrees, as indicated in the basic equation
(\ref{2.3}). The theory of multivectors, based on Clifford algebra, was
developed by Hestenes ($^{14}$). In the following  some useful 
formulas are displayed without proofs.

For a vector $a$ and an $r$-vector $A_r$ the inner and the
outer product are defined according to
\be
   a \cdot A_r \equiv {1\oo 2} \left ( a A_r - (-1)^r A_r a \right ) =
   - (-1)^r A_r \cdot a
\lbl{3.13a}
\ee
\be
      a \wedge A_r = {1 \oo 2} \left ( a A_r + (-1)^r A_r a \right )  =
      (-1)^r A_r \wedge a 
\lbl{3.13b}
\ee
The inner product has symmetry opposite to that of the outer product,
therefore the signs in front of the second terms in the above equations
are different.      

Combining (\ref{3.13a}) and (\ref{3.13b}) we find
\be
    a A_r = a \cdot A_r + a \wedge A_r
\lbl{3.13g}
\ee
For $A_r = a_1 \wedge a_2 \wedge ... \wedge a_r$ eq.(\ref{3.13a}) can be
evaluated to give the useful expansion
\be
   a \cdot (a_1 \wedge ... \wedge a_r) = \sum_{k=1}^r (-1)^{k+1}(a \cdot a_k)
   a_1 \wedge ... a_{k-1} \wedge a_{k+1} \wedge ... a_r
\lbl{3.13h}
\ee
In particular,
\be
      a \cdot (b \wedge c) = (a \cdot b)c - (a \cdot c) b
\lbl{3.13i}
\ee        

It is very convenient to introduce, besides the basis vectors $e_{\mu}$,
another set of basis vectors $e^{\nu}$ by the condition
\be
    e_{\mu} \cdot e^{\nu} = {\delta_{\mu}}^{\nu}
\lbl{ 3.13c}
\ee
Each $e^{\mu}$ is a linear combination of $e_{\nu}$:
\be
     e^{\mu} = g^{\mu \nu} e_{\nu}
\lbl{3.13d}
\ee
from which we have
\be
    g^{\mu \alpha} g_{\alpha \nu} = {\delta_{\mu}}^{\nu}
\lbl{3.13e}
\ee
and 
\be
     g^{\mu \nu} = e^{\mu} \cdot e^{\nu} = {1\oo 2} (e^{\mu \nu} +
     e^{\nu} e^{\mu} )
\lbl{3.13f}
\ee         

Let $e_1, \, e_2, \, ..., \, e_n$ be linearly independent vectors,
and $\alpha, \, \alpha^i, \, \alpha^{i_1 i_2}, ...$ scalar coefficients.
A generic Clifford number can then be written as
\be
       A = \alpha + \alpha^i e_i + {1\oo {2!}} \, \alpha^{i_1 i_2}
       \,e_{i_1} \wedge e_{i_2} + ... {1\oo {n!}} \, \alpha^{i_1 ... i_n}
       e_{i_1} \wedge ... \wedge e_{i_n}
\lbl{3.14}
\ee

Since it is a superposition of multivectors of all possible grades
it will be called {\it polyvector}.\footnote{
Following a suggestion by Pezzaglia ($^{10}$) I call a generic 
Clifford number
{\it polyvector} and reserve the name {\it multivector} for an $r$-vector,
since the latter name is already widely used for the corresponding object
in the calculus of differential forms.} Another name, also often used
in the literature, is {\it Clifford aggregate}. These mathematical objects
have far reaching geometrical and physical implications that will be
discussed and explored to some extent in the rest of the paper.

\subsection{The algebra of spacetime}

In spacetime we have 4 linearly independent vectors $e_{\mu}$, $\mu =
0,1,2,3$. Let us consider {\it flat} spacetime. It is convenient then to
take orthonormal basis vectors $\gamma_{\mu}$
\be
       \gamma_{\mu} \cdot \gamma_{\nu} = \eta_{\mu \nu}
\lbl{3.14a}
\ee
where $\eta_{\mu \nu}$ is the diagonal metric tensor with signature
$(+ - - -)$. 

The Clifford algebra in $V_4$ is called the {\it Dirac algebra}. Writing
$\gamma_{\mu \nu} \equiv \gamma_{\mu} \wedge \gamma_{\nu}$ for a basis
bivector, $\gamma_{\mu \nu \rho} \equiv  \gamma_{\mu} \wedge \gamma_{\nu} 
\wedge \gamma_{\rho}$ for a basis trivector and $\gamma_{\mu \nu \rho
\sigma} \equiv  \gamma_{\mu} \wedge \gamma_{\nu} \wedge \gamma_{\rho} 
\wedge \gamma_{\sigma}$ for a basis quadrivector we can express
an arbitrary number of Dirac algebra as
\be
     D = \sum_r D_r = d + d^{\mu} \gamma_{\mu} + {1 \oo {2!}} \, d^{\mu \nu}
     \gamma_{\mu \nu} + {1\oo {3!}} \, d^{\mu \nu \rho} \gamma_{\mu \nu \rho} +
     {1\oo {4!}} \, d^{\mu \nu \rho \sigma} \gamma_{\mu \nu \rho \sigma}
\lbl{3.15}
\ee
where $d, \, d^{\mu}, \, d^{\mu \nu}, ...$ are scalar coefficients.

Let us introduce
\be
   \gamma_5 \equiv \gamma_0 \wedge \gamma_1 \wedge \gamma_2 \wedge \gamma_3
   = \gamma_0 \gamma_1 \gamma_2 \gamma_3 \; \; , \qquad \gamma_5^2 = - 1
\lbl{3.16}
\ee
which is the unit element of 4-dimensional volume and is called
{\it pseudoscalar}. Using the relations
\be
    \gamma_{\mu \nu \rho \sigma} = \gamma_5 \epsilon_{\mu \nu \rho \sigma}
\lbl{3.17}
\ee
\be
    \gamma_{\mu \nu \rho} = \gamma_{\mu \nu \rho \sigma} \gamma^{\rho}
\lbl{3.18}
\ee
where  $\epsilon_{\mu \nu \rho \sigma}$ is the totally
antisymmetric tensor  and introducing the new coefficients
  $$ S \equiv d \; , \quad V^{\mu} \equiv d^{\mu} \; , \quad T^{\mu \nu}
  \equiv{1\oo 2} d^{\mu \nu}$$
\be
    C_{\sigma} \equiv {1\oo {3!}} d^{\mu \nu \rho} 
    \epsilon_{\mu \nu \rho \sigma} \; , \qquad P \equiv {1\oo {4!}}
    d^{\mu \nu \rho \sigma} \epsilon_{\mu \nu \rho \sigma} 
\lbl{3.19}
\ee
we can rewrite $D$ of eq.(\ref{3.15}) as the sum of scalar, vector,
bivector, pseudovector and pseudoscalar part:
\be
    D = S + V^{\mu} \gamma_{\mu} + T^{\mu \nu} \gamma_{\mu \nu} + 
    C^{\mu} \gamma_5 \gamma_{\mu} + P \gamma_5
\lbl{3.20}
\ee

\subsection{Polyvector fields}

A polyvector may depend on spacetime points. Let $A = A(x)$ be an $r$-vector
field. Then one can define the {\it gradient operator} according to
\be
     \p = \gamma^{\mu} \p_{\mu}
\lbl{3.21}
\ee
where $\p_{\mu}$ is the usual partial derivative. The gradient
operator $\p$ can act on any $r$-vector field. Using (\ref{3.13g}) we
have
\be
      \p A = \p \cdot A + \p \wedge A
\lbl{3.22}
\ee

{\it Example}. Let $A = a = a_{\nu} \gamma^{\nu}$ be a 1-vector field.
Then
\begin{eqnarray}
    \p a &=& \gamma^{\mu} \p_{\mu} (a_{\nu} \gamma^{\nu}) = \gamma^{\mu}
    \cdot \gamma^{\nu} \, \p_{\mu} a^{\nu} + \gamma^{\mu} \wedge \gamma^{\nu}
    \p_{\mu} a_{\nu} \nonumber \\
    &=& \p_{\mu} a^{\mu} + {1\oo 2} (\p_{\mu} a_{\nu} - \p_{\nu} a_{\mu})
    \gamma^{\mu} \wedge \gamma^{\nu}
\lbl{3.23}
\end{eqnarray}
The simple expression $\p a$ thus contains a scalar and bivector part,
the former being the usual divergence and the latter the usual curl of
a vector field.

\paragraph{Maxwell's equations}  We shall demonstrate now by a concrete physical
example the usefulness of Clifford algebra. Let us consider the electromagnetic
field which, in the language of Clifford algebra, is a bivector field $F$.
The source of the field is the electromagnetic current $j$ which is a 1-vector
field. Maxwell's equations read
\be 
    \p F = 4 \pi j
\lbl{3.24}
\ee
The grade of the gradient operator $\p$ is 1. Therefore we can use
the relation (\ref{3.22}) and we find that eq.(\ref{3.24}) becomes
\be
    \p \cdot F + \p \wedge F = 4 \pi j
\lbl{3.25}
\ee
which is equivalent to
\be
    \p \cdot F = - 4 \pi j
\lbl{3.25a}
\ee
\be
     \p \wedge F = 0
\lbl{3.25b}
\ee
since the first term on the left of eq.(\ref{3.25}) is a vector and the second
term is a bivector. This results from the general relation (\ref{3.25} ).
It can also be explicitly demonstrated. Expanding
\be
    F = {1\oo 2} F^{\mu \nu} \, \gamma_{\mu} \wedge \gamma_{\nu}
\lbl{3.26}
\ee
\be
   j = j^{\mu} \gamma_{\mu}
\lbl{3.27a}
\ee
we have
\begin{eqnarray}
      \p \cdot F & = & \gamma^{\alpha} \p_{\alpha} \cdot ({1\oo 2} F^{\mu \nu}
      \gamma_{\mu} \wedge \gamma_{\nu}) = {1\oo 2} \gamma^{\alpha} \cdot
      (\gamma_{\mu} \wedge \gamma_{\nu}) \p_{\alpha} F^{\mu \nu} \nonumber \\
       & = & {1\oo 2} \left ( (\gamma^{\alpha} \cdot \gamma_{\mu}) \gamma_{\nu} -
       (\gamma^{\alpha} \cdot \gamma_{\nu})\gamma_{\mu} \right ) \p_{\alpha}
       F^{\mu \nu} = \p_{\mu} F^{\mu \nu} \, \gamma_{\nu}
\lbl{3.27}
\end{eqnarray}
\be
     \p \wedge F = {1\oo 2} \gamma^{\alpha} \wedge \gamma_{\mu} \wedge
     \gamma_{\nu} \, \p_{\alpha} F^{\mu \nu} = {1\oo 2} {\epsilon^{\alpha}}_{
     \mu \nu \rho} \, \p_{\alpha} F^{\mu \nu} \gamma_5 \gamma^{\rho}
\lbl{3.28}
\ee
where we have used (\ref{3.13i} ) and eqs.(\ref{3.17}),(\ref{3.18}).
From the above considerations it then follows that the compact equation 
(\ref{3.24}) is equivalent to the usual tensor form of Maxwell's equations
\be
    \p_{\nu} F^{\mu \nu} = - 4 \pi j^{\mu}
\lbl{3.29}
\ee
\be
    {\epsilon^{\alpha}}_{\mu \nu \rho} \, \p_{\alpha} F^{\mu \nu} = 0
\lbl{3.30}
\ee
Applying the gradient operator $\p$ to the left and to the right side
of eq.(\ref{3.24}) we have
\be
     \p^2 F = \p j
\lbl{3.31}                
\ee
Since $\p^2 = \p \cdot \p + \p  \wedge \p = \p \cdot \p$ is a scalar
operator, $\p^2 F$ is a bivector. The right hand side of eq.(\ref{3.31}) gives
\be
    \p j = \p \cdot j + \p \wedge j
\lbl{3.32}
\ee

Equating the terms of the same grade on the left and the right hand side of
eq.(\ref{3.31}) we obtain
\be
     \p^2 F = \p \wedge j
\lbl{3.33}
\ee
\be
     \p \cdot j = 0
\lbl{3.34}
\ee
The last equation expresses the conservation of the electromagnetic current.

\paragraph{Motion of a charged particle} In this example we wish to go a step
forward. Our aim is not only to describe how a charged particle moves in an
electromagnetic field, but also include particle's (classical) spin. Therefore,
following Pezzaglia ($^{10}$), we define the 
{\it momentum polyvector} $P$ as the
{\it vector momentum} $p$ plus the bivector {\it spin angular momentum}
$S$
\be
    P = p + S
\lbl{3.35}
\ee
or in components
\be
    P = p^{\mu} \gamma_{\mu} + {1\oo 2} S^{\mu \nu} \, \gamma_{\mu} \wedge
    \gamma_{\nu}
\lbl{3.36}
\ee
We also assume that the condition $p_{\mu} S^{\mu \nu} = 0$ is
satisfied. The latter condition ensures the spin to be a simple bivector,
which is purely spacelike in the rest frame of the particle. The polyvector
equation of motion is
\be {\dot P} \equiv {{{\dd} P} \oo {{\dd} \tau}} = {e \oo {2m}} \, [P, F]
\lbl{3.37}
\ee
where $[P,F] \equiv PF - FP$. The vector and bivector parts of 
eq.(\ref{3.37}) are
\be
     {\dot p}^{\mu} = {e\oo m} \, {F^{\mu}}_{\nu} p^{\nu}
\lbl{3.38}
\ee
\be
     {\dot S}^{\mu \nu} = {e\oo {2m}} ({F^{\mu}}_{\alpha} S^{\alpha \nu}
     - {F^{\nu}}_{\alpha} S^{\alpha \mu})
\lbl{3.39}
\ee
These are just the equation of motion for linear momentum and spin,
respectively.

\subsection{Physical quantities as polyvectors}

The compact equations at the end of the last subsection suggest
a generalization that every physical quantity is a polyvector. We shall
explore such an assumption and see how far we can get.

In 4-dimensional spacetime {\it the momentum polyvector} is
\be
    P = \mu + p^{\mu} e_{\mu} + S^{\mu \nu} e_{\mu} e_{\nu} +
    \pi^{\mu} e_5 e_{\mu} + m e_5
\lbl{3.40}
\ee
and {\it the velocity polyvector} is
\be
   {\dot X} = {\dot \sigma} + {\dot x}^{\mu} e_{\mu} + {\dot \alpha}^{\mu \nu}
   e_{\mu} e_{\nu} + {\dot \xi}^{\mu} e_5 e_{\mu} + {\dot s} e_5
\lbl{3.41}
\ee
where $e_{\mu}$ are four basis vectors satisfying
\be
       e_{\mu} \cdot e_{\nu} = \eta_{\mu \nu}
\lbl{3.41a}
\ee
and $e_5 \ \equiv e_0 e_1 e_2 e_3$ is the pseudoscalar. For the purposes
which will become clear later we now use the symbols $e_{\mu}$, $e_5$
instead of $\gamma_{\mu}$ and $\gamma_5$.

We associate with each particle the velocity polyvector ${\dot X}$ and
its conjugate momentum polyvector $P$. These quantities are  generalizations
of the point particle 4-velocity ${\dot x}$ and its conjugate momentum $p$.
Besides a vector part we now include the scalar part ${\dot \sigma}$, the
bivector part ${\dot \alpha}^{\mu \nu} e_{\mu} e_{\nu}$, the pseudovector
part ${\dot \xi}^{\mu} e_5 e_{\mu}$ and the pseudoscalar part ${\dot s} e_5$
into the definition of a particle's velocity, and analogously for a particle's
momentum. We would like now to derive the equations of motion which will
tell us how those quantities depend on the evolution parameter $\tau$. For
simplicity we consider a free particle.

Let the action be a straightforward generalization of the first order or
phase space action of the usual constrained point particle
relativistic theory:
\be
    I[X,P,\lambda] = {1 \oo 2} \int {\dd} \tau \, \left ( P {\dot X} + {\dot X} P
    - \lambda P^2 \right )
\lbl{3.42}
\ee
where $\lambda$ is a scalar Lagrange multiplier. Variation of (\ref{3.42}) with
respect to $\lambda$ gives the constraint
\be
       P^2 = 0
\lbl{2.6}
\ee
Using the definition (\ref{3.40}), the last equation becomes\footnote{For
more details see ($^{13}$).}
\be
      P^2 = p^2 - m^2 - \pi^2 + \mu^2 + 2 \mu (p^{\mu} e_{\mu} + m e_5) + \;
      {\rm etc.} = 0
\lbl{2.7}
\ee

After quantization the above constraint becomes
\be
    {\hat P}^2 \Phi = 0
\lbl{2.8}
\ee
where $\Phi$ is a polyvector valued wave function, or briefly,
{\it polyvector wave function} ($^{13}$).

A particular class of solutions satisfies
\be
     {\hat P} \Psi = 0
\lbl{2.9}
\ee

In particular, when the state represented by $\Psi$ has definite values
$\mu = 0$, $S^{\mu \nu} = 0$, $\pi^{\mu} = 0$, then
\be
    {\hat P} \Psi = ({\hat p}^{\mu} e_{\mu} + m e_5) \Psi = 0
\lbl{2.10}
\ee
or
\be
      ({\hat p}^{\mu} \gamma_{\mu} - m) \Psi = 0
\lbl{2.11}
\ee
where
\be
     \gamma_{\mu} \equiv e_5 e_{\mu} \nonumber
\ee
\be
    \gamma_5 \equiv \gamma_0 \gamma_1 \gamma_2 \gamma_3 =
    e_0 e_1 e_2 e_3 \equiv e_5
\lbl{2.12}
\ee

We have thus found that the {\it Dirac equation} (\ref{2.11})
is a special case of the equation (\ref{2.9}), which in turn is the
``square root" of the generalized Klein-Gordon equation (\ref{2.8}).
The latter equation involves the polyvector wave function $\Psi$.
Amongst various possible polyvector wave functions there are such
that satisfy eq.(\ref{2.11}), i.e., the Dirac equation. The 
latter equation describes a spin $1\oo 2$ particle, and $\Psi$ satisfying 
(\ref{2.11}) is a spinor. This obviously means that spinors can be
represented as a sort of polyvectors.

We have thus arrived at a very important observation, namely that a
generic polyvector contains spinors. A generic polyvector wave function
contains bosons and fermions.

To illustrate this let us consider the 3-dimensional space $V_3$. Basis
vectors are $\sigma_1$, $\sigma_2$, $\sigma_3$ and they satisfy the
Pauli algebra
\be
     \sigma_i \cdot \sigma_j \equiv {1\oo 2} (\sigma_i \sigma_j +
     \sigma_j \sigma_i) = \delta_{i j} \; , \qquad i , j = 1, 2, 3
\lbl{3.124}
\ee

The unit pseudoscalar
\be
    \sigma_1 \sigma_2 \sigma_3 \equiv I
\lbl{3.125}
\ee
commutes with all elements of the Pauli algebra and its square is
$I^2 = -1$. It behaves as the ordinary imaginary unit $i$. Therefore,
in 3-space, we may identify the imaginary unit $i$ with the unit pseudoscalar
$I$.

An arbitrary polyvector in $V_3$ can be written in the form\footnote{
Here I review and adapt the Hestenes procedure ($^{14}$).}
\be
     \Phi = \alpha^0 + \alpha^i \sigma_i + i \beta^i \sigma_i + i \beta
     = \Phi^0 + \Phi^i \sigma_i
\lbl{3.126}
\ee
where $\Phi^0$, $\Phi^i$ are formally complex numbers.

We can decompose ($^{14}$):
\be
    \Phi = \Phi {1\oo 2} (1 + \sigma_3) + \Phi {1\oo 2} (1 - \sigma_3) =
    \Phi_+ + \Phi_-
\lbl{3.127}
\ee
where $\Phi \in {\cal I}_+$ and $\Phi_- \in {\cal I}_-$ are
independent minimal {\it left ideals}.

Let us recall the definition of ideal.
A left ideal ${\cal I}_L$ in an algebra $C$ is a set of elements such
that if $a \in {\cal I}_L$ and $c \in C$, then $c a \in {\cal I}_L$.
If $a \in {\cal I}_L$, $b \in {\cal I}_L$, then $(a + b) \in {\cal I}_L$.
A right ideal ${\cal I}_R$ is defined similarly except that $a c \in {\cal I}_R$.
A left (right) minimal ideal is a left (right) ideal which contains no
other ideals but itself and the null ideal.

A basis in ${\cal I}_+$ is given by two polyvectors
\be
       u_1 = {1\oo 2} (1 + \sigma_3) \; \; , \quad u_2 = (1 - \sigma_3) \sigma_1
\lbl{3.128}
\ee
which satisfy
\begin{eqnarray}
    \sigma_3 u_1 &=& \ u_1 \quad \sigma_1 u_1 = u_2 \quad \sigma_2 u_1 = 
    \ i u_2 \nonumber \\
    \sigma_3 u_2 &=& - u_2 \quad \sigma_1 u_2 = u_1 \quad
    \sigma_2 u_2 = - i u_1
\lbl{3.129}
\end{eqnarray}
These are precisely the well known relations for basis spinors. Thus
we have arrived at the very profound result that the polyvectors $u_1$, $u_2$
behave as basis spinors.

Similarly, a basis in ${\cal I}_+$ is given by
\be
    v_1 = {1\oo 2} (1 + \sigma_3) \sigma_1 \; \; , \quad 
    v_2 = {1\oo 2} (1 - \sigma_3)
\lbl{3.130}
\ee
and satisfies    
\begin{eqnarray}
    \sigma_3 v_1 &=& \ v_1 \quad \sigma_1 v_1 = v_2 \quad \sigma_2 v_1 = 
    \ i v_2 \nonumber \\
    \sigma_3 v_2 &=& - v_2 \quad \sigma_1 v_2 = v_1 \quad
    \sigma_2 v_2 = - i v_1
\lbl{3.131}
\end{eqnarray}

A polyvector $\Phi$ can be written in {\it spinor basis}
\be
    \Phi = \Phi_+^1 u_1 + \Phi_+^2 u_2 + \Phi_-^1 v_1 + \Phi_-^2 v_2
\lbl{3.132}
\ee
where
\begin{eqnarray}
   \Phi_+^1 &=& \Phi^0 + \Phi^3  \; \; , \quad \Phi_-^1 = \Phi^1 - i \Phi^2
   \nonumber \\
   \Phi_+^2 &=& \Phi^1 + i \Phi^2 \; \; , \quad \Phi_-^2 = \Phi^0 - \Phi^3
\lbl{3.133}
\end{eqnarray}
Eq.(\ref{3.132}) is an alternative expansion of a polyvector. We can
expand the same polyvector $\Phi$ either according to (\ref{3.126}) or
according to (\ref{3.132}).

Introducing the matrices
\be
     \xi_{a b} = \pmatrix{
     u_1 & v_1 \cr
     u_2 & v_2 \cr } \; \; \quad \Phi^{a b} = \pmatrix{
     \Phi_+^1 & \Phi_-^1 \cr
     \Phi_+^2 & \Phi_-^2 \cr }
\lbl{3.134}
\ee
we can write (\ref{3.132}) as
\be
      \Phi = \Phi^{a b} \xi_{a b}
\lbl{3.135}
\ee

Thus a polyvector can be represented as a {\it matrix} $\Phi^{a b}$. The
decomposition (\ref{3.127}) then reads
\be
    \Phi = \Phi_+ + \Phi_- = (\Phi_+^{a b} + \Phi_-^{a b}) \xi_{a b}
\lbl{3.136}
\ee
where
\be
  \Phi_+^{a b} = \pmatrix{
         \Phi_+^1 & 0 \cr
         \Phi_+^2 & 0 \cr }
\lbl{3.137}
\ee
\be
     \Phi_-^{a b} = \pmatrix{
     0 & \Phi_-^1 \cr
     0 & \Phi_-^2 \cr }
\lbl{3.138}
\ee

From (\ref{3.135}) we can directly calculate the matrix elements $\Phi^{ab}$.
We only need to introduce the new elements $\xi^{\dagger ab}$ which
satisfy
\be
     ({\xi^{\dagger}}^{a b} \xi_{c d} )_S = {\delta^a}_c {\delta^b}_d
\lbl{3.139}
\ee
The superscript $\dagger$ means Hermitian conjugation ($^{14}$). If
\be
      A = A_S + A_V +A_B + A_P
\lbl{3.139a}
\ee
is a Pauli number, then
\be
     A^{\dagger} = A_S + A_V - A_B - A_P
\lbl{3.139b}
\ee
This means that the order of basis vectors $\sigma_i$ in the expansion
of $A^{\dagger}$ is reversed. Thus $u_1^{\dagger} = u_1$, but
$u_2^{\dagger} = {1\oo 2} (1 + \sigma_3) \sigma_1$. Since $(u_1^{\dagger}
u_1)_S = {1\oo 2}$, $(u_2^{\dagger} u_2)_S = {1\oo 2}$ it is convenient
to introduce ${u^{\dagger}}^{1} = 2 u_1$ and ${u^{\dagger}}^2 = 2 u_2$
so that $({u^{\dagger}}^1 u_1)_S = 1$, $({u^{\dagger}}^2 u_2)_S = 1$. If we
define similar relations for $v_1$, $v_2$ then we obtain (\ref{3.139}).

From (\ref{3.135}) and (\ref{3.139}) we have
\be
     \Phi^{a b} = ({\xi^{\dagger}}^{a b} \Phi )_I
\lbl{3.140}
\ee
Here the subscript $I$ means {\it invariant part}, i.e. scalar plus
pseudoscalar part (remember that pseudoscalar unit has here the role of
imaginary unit and that $\Phi^{a b}$ are thus complex numbers).

The relation (\ref{3.140}) tells us how from an arbitrary polyvector $\Phi$
(i.e. a Clifford number) can we obtain its {\it matrix representation}
$\Phi^{a b}$.

$\Phi$ in (\ref{3.140}) is an arbitrary Clifford number. In particular $\Phi$
may be any of the basis vectors $\sigma_i$.

{\it Example}  $\Phi = \sigma_1$:
\begin{eqnarray}
     \Phi^{11} &=& ({\xi^{\dagger}}^{11} \sigma_1)_I = ({u^{\dagger}}^1
     \sigma_1)_I = \left ( (1 + \sigma_3)\sigma_1 \right )_I = 0 \nonumber \\
     \Phi^{12}  &=& ({\xi^{\dagger}}^{12} \sigma_1)_I = ({v^{\dagger}}^1
     \sigma_1)_I = \left ( (1 - \sigma_3)\sigma_1 \sigma_1 \right )_I = 
     1 \nonumber \\
    \Phi^{21}  &=& ({\xi^{\dagger}}^{21} \sigma_1)_I = ({u^{\dagger}}^2
     \sigma_1)_I = \left ( (1 + \sigma_3)\sigma_1 \sigma_1 \right )_I = 
     1 \nonumber \\
    \Phi^{22}  &=& ({\xi^{\dagger}}^{22} \sigma_1)_I = ({v^{\dagger}}^2
     \sigma_1)_I = \left ( (1 - \sigma_3)\sigma_1 \right )_I = 0 
\lbl{3.141}
\end{eqnarray}
Therefore
\be
     (\sigma_1)^{a b} = \pmatrix{
     0 & 1 \cr
     1 & 0 \cr }
\lbl{3.142}
\ee
Similarly we obtain from (\ref{3.140}) when $\Phi = \sigma_2$ and
$\Phi = \sigma_3$, respectively, that
\be
     (\sigma_2)^{a b} = \pmatrix{
     0 & -i \cr
     i & 0 \cr }  \; \; , \qquad
     (\sigma_3)^{a b} = \pmatrix{
     1 & 0 \cr
     0 & -1 \cr }
\lbl{3.143}
\ee

So we have obtained the matrix representation of the basis vectors $\sigma_i$.
Actually (\ref{3.142}),(\ref{3.143}) are the well known {\it Pauli matrices}.

When $\Phi = u_1$ and $\Phi = u_2$, respectively, we obtain
\be
     (u_1)^{a b} = \pmatrix{
     1 & 0 \cr
     0 & 0 \cr }  \; \; , \qquad
     (u_2)^{a b} = \pmatrix{
     0 & 0 \cr
     1 & 0 \cr }
\lbl{3.144}
\ee
which are a matrix representation of the {\it basis spinors} $u_1$ and $u_2$.

Similarly we find
\be
     (v_1)^{a b} = \pmatrix{
     0 & 1 \cr
     0 & 0 \cr }  \; \; , \qquad
     (v_2)^{a b} = \pmatrix{
     0 & 0 \cr
     0 & 1 \cr }
\lbl{3.145}
\ee

In general a {\it spinor} is a superposition
\be
     \psi = \psi^1 u_1 + \psi^2 u_2
\lbl{3.146}
\ee
and its matrix representation is
\be
       \psi \rightarrow \pmatrix{
       \psi^1 & 0 \cr
       \psi^2 & 0 \cr}
\lbl{3.147}
\ee

Another independent spinor is
\be
     \chi = \chi^1 v_1 + \chi^2 v_2
\lbl{3.148}
\ee
with matrix representation
\be
     \chi \rightarrow \pmatrix{
     0 & \chi^1 \cr
     0 & \chi^2 \cr}
\lbl{3.149}
\ee

If we multiply a spinor $\psi$ from the left by any element $R$ of the
Pauli algebra we obtain another spinor
\be
     \psi' = R \psi \rightarrow \pmatrix{
     \psi'^1 & 0 \cr
     \psi'^2 & 0\cr}
\lbl{3.150}
\ee
which is an element of the same minimal left ideal. Therefore,
if only multiplication from the left is considered, a spinor can be
considered as a column matrix
\be
     \psi \rightarrow \pmatrix{
     \psi^1 \cr
     \psi^2 \cr}
\lbl{3.151}
\ee
This is just the common representation of spinors. But it is not
general enough to be valid for all the interesting situations which
occur in the Clifford algebra.

We have thus arrived at a very important finding. {\it Spinors} are just
particular {\it Clifford numbers}: they belong to a left or right minimal
ideal. For instance, a generic {\it spinor} is
\be
     \psi = \psi^1 u_1 + \psi^2 u_2  \quad {\rm with} \quad
     \Phi^{ab} = \pmatrix{
     \psi^1 & 0 \cr
     \psi^2 & 0 \cr}
\lbl{3.152}
\ee
A {\it conjugate spinor} is
\be
     \psi^{\dagger} = \psi^{1*} u_1^{\dagger} + \psi^{2*} u_2^{\dagger}
     \quad {\rm with} \quad
     {(\Phi^{ab})}^* = \pmatrix{
     {\psi^1}^* & {\psi^2}^* \cr
     0 & 0 \cr}
\lbl{3.153}
\ee
and it is an element of a minimal {\it right ideal}.

The above consideration can be generalized to 4 or more dimensions (see
($^{15}$)).

Scalars, vectors, etc., and spinors can be reshuffled by the elements
of Clifford algebra. For instance, vectors can be transformed into spinors,
and vice verse. Within Clifford algebra we have thus transformations 
which change bosons into fermions! It remains to be investigated whether
such a kind of ``supersymmetry" is related to the well known supersymmetry.

\subsection{Relativity of signature}

In the previous subsection we have seen how Clifford algebra can be used 
in the formulation of the point particle classical and quantum theory.
The metric of spacetime was assumed as usually to have the Minkowski signature,
and we have used the choice $(+ - - -)$.
We are now going to find out that within
Clifford algebra the signature is a matter of choice of basis vectors amongst
the available Clifford numbers.

Suppose we have a 4-dimensional space $V_4$ with signature (+ + + +). Let
$e_{\mu},\, \mu = 0,1,2,3$ be basis vectors satisfying
\be
     e_{\mu} \cdot e_{\nu} \equiv {1\oo 2} (e_{\mu} e_{\nu} + e_{\nu} e_{\mu})
     = \delta_{\mu \nu}
\lbl{4.1}
\ee
where $\delta_{\mu \nu}$ is the {\it Euclidean signature} of $V_4$.
The vectors $e_{\mu}$ can be used as generators of Clifford algebra
${\cal C}$ over $V_4$ with a generic Clifford number (named also
polyvector or Clifford aggregate) expanded in term of $e_J = (1,e_{\mu},e_{\mu
\nu}, e_{\mu \nu \alpha}, e_{\mu \nu \alpha, \beta})$,
$\mu < \nu < \alpha < \beta$,
\be
        A = a^J e_J = a + a^{\mu} e_{\mu} + a^{\mu \nu} e_{\mu} e_{\nu} +
        a^{\mu \nu \alpha} e_{\mu} e_{\nu} e_{\alpha} + a^{\mu \nu \alpha \beta}
        e_{\mu} e_{\nu} e_{\alpha} e_{\beta}
\lbl{4.2}
\ee
Let us consider the set of four Clifford numbers $(e_0,\, e_i e_0)$,
$i = 1,2,3$ and denote them as
\begin{eqnarray}
         e_0 &\equiv& \gamma_0        \nonumber \\
         e_i e_0 &\equiv& \gamma_i
\lbl{4.3}
\end{eqnarray}
The Clifford numbers $\gamma_{\mu}$, $\mu = 0,1,2,3$ satisfy
\be
         {1\oo 2} (\gamma_{\mu} \gamma_{\nu} + \gamma_{\nu} \gamma_{\mu})
         = \eta_{\mu \nu}
\lbl{4.4}
\ee
where $\eta_{\mu \nu} = {\rm diag} (1, -1, -1 ,-1)$ is the {\it Minkowski
tensor}. We see that the $\gamma_{\mu}$ behave as basis vectors in a
4-dimensional space $V_{1,3}$ with signature $(+ - - -)$. We can form a
Clifford aggregate
\be
     \alpha = \alpha^{\mu} \gamma_{\mu}
\lbl{4.5}
\ee
which has the properties of a {\it vector} in $V_{1,3}$. From the point of
view of the space $V_4$ the same object $\alpha$ is a linear combination of
a vector and bivector:
\be
       \alpha = \alpha^0 e_0 + \alpha^i e_i e_0
\lbl{4.6}
\ee
We may use $\gamma_{\mu}$ as generators of the Clifford algebra
${\cal C}_{1,3}$ defined over the pseudo-Euclidean space $V_{1,3}$. The
basis elements of ${\cal C}_{1,3}$ are $\gamma_J =$\ $(1,\gamma_{\mu},\gamma_{\mu
\nu}, \gamma_{\mu \nu \alpha}, \gamma_{\mu \nu \alpha \beta})$, with
$\mu < \nu < \alpha < \beta$. A generic  Clifford aggregate in ${\cal C}_{1,3}$
is given by
\be
        B = b^J \gamma_J = b + b^{\mu} \gamma_{\mu} + 
        b^{\mu \nu} \gamma_{\mu} \gamma_{\nu} +
        b^{\mu \nu \alpha} \gamma_{\mu} \gamma{\nu} \gamma_{\alpha} + 
        b^{\mu \nu \alpha \beta}
        \gamma_{\mu} \gamma_{\nu} \gamma_{\alpha} \gamma_{\beta}
\lbl{4.7}
\ee
With suitable choice of the coefficients $b^J = (b, b^{\mu},
b^{\mu \nu}, b^{\mu \nu \alpha}, b^{\mu \nu \alpha, \beta})$
we have that $B$ of eq.(\ref{4.7}) is equal to $A$ of e.(\ref{4.2}). Thus the
same number $A$ can be described either within ${\cal C}_4$ or within
${\cal C}_{1,3}$. The expansions (\ref{4.7}) and (\ref{4.2}) 
exhaust all possible
numbers of the Clifford algebras ${\cal C}_{1,3}$ and ${\cal C}_4$.
The algebra ${\cal C}_{1,3}$ is isomorphic to the
algebra ${\cal C}_4$ and actually they are just two different representations
of the same set of Clifford numbers (called also polyvectors or Clifford
aggregates).

As an alternative to (\ref{4.3}) we can choose
\begin{eqnarray}
          e_0 e_3 &\equiv& {\tilde \gamma}_0 \nonumber \\
          e_i &\equiv& {\tilde \gamma}_i
\lbl{4.8}
\end{eqnarray}
from which we have
\be
      {1\oo 2} ({\tilde \gamma}_{\mu} {\tilde \gamma}_{\nu} + {\tilde 
      \gamma}_{\nu} {\tilde \gamma}_{\mu} ) = {\tilde \eta}_{\mu \nu}
\lbl{4.9}
\ee
with ${\tilde \eta}_{\mu \nu} = {\rm diag} (- 1, 1, 1, 1)$. Obviously
${\tilde \gamma}_{\mu}$ are basis vectors of a pseudo-Euclidean space
${\tilde V}_{1,3}$ and they generate the Clifford algebra over 
${\tilde V}_{1,3}$ which is
yet another representation of the same set of objects (i.e. polyvectors). But
the spaces $V_4$, $V_{1,3}$ and ${\tilde V}_{1,3}$ are not the same and
they span different subsets of polyvectors. In a similar way we can obtain
spaces with signatures $(+ - + +)$, $(+ + - +)$, $(+ + + -)$, 
$(- + - -)$, $(- - + -)$,
$(- - - +)$ and corresponding higher dimensional analogs. But we cannot obtain 
signatures of the type $(+ + - -)$, $(+ - + -)$, etc. In order to obtain such
signatures we proceed as follows.

{\it 4-space.} \hs{3mm}
First we observe that the bivector ${\bar I} = e_3 e_4$ satisfies ${\bar I}^2
= -1$, commutes with $e_1,\, e_2$ and anticommutes with $e_3,\, e_4$. So we
obtain that the set of Clifford numbers $\gamma_{\mu} = (e_1 {\bar I}, e_2
{\bar I}, e_3, e_3)$ satisfies
\be
      \gamma_{\mu} \cdot \gamma_{\nu}  = {\bar \eta}_{\mu \nu}
\lbl{4.9a}
\ee
where ${\bar \eta} = {\rm diag} ( -1,-1,1,1)$.

{\it 8-space.} \hs{3mm} Let $e_A$ be basis vectors of 8-dimensional\footnote{
The Clifford Algebra of 8-dimensional space was studied in ref. ($^{16}$),
where it was shown that octonions are imbedded in ${\cal C}_8$.} 
vector space with signature (+ + + + + + + +). Let us decompose
\begin{eqnarray}
      e_A = (e_{\mu}, e_{\bar \mu}) \; \; \qquad \mu &=& 0,1,2,3 \nonumber \\
      {\bar \mu} &=& {\bar 0}, {\bar 1}, {\bar 2}, {\bar 3}
\lbl{4.10}
\end{eqnarray}
The inner product of two basis vectors
\be
      e_A \cdot e_B = \delta_{AB}     \nonumber
\ee
then splits into the following set of equations:
\begin{eqnarray}
     e_{\mu} \cdot e_{\nu} &=& \delta_{\mu \nu}  \nonumber \\
     e_{\bar \mu} \cdot e_{\bar \nu} &=& \delta_{{\bar \mu} {\bar \nu}} 
     \nonumber \\
     e_{\mu} \cdot e_{\bar \nu} &=& 0
\lbl{4.11}
\end{eqnarray}
The number ${\bar I} = e_{\bar 0} e_{\bar 1} e_{\bar 2} e_{\bar 3}$ has
the propoerties
\begin{eqnarray}
      {\bar I}^2 &=& 1 \nonumber \\
      {\bar I} e_{\mu} &=& e_{\mu} {\bar I} \nonumber \\
      {\bar I} e_{\bar \mu} &=& - e_{\bar \mu} {\bar I}
\lbl{4.12}
\end{eqnarray}
The set of numbers
\begin{eqnarray}
       \gamma_{\mu} &=& e_{\mu}    \nonumber \\
       \gamma_{\bar \mu} &=& e_{\bar \mu} {\bar I}
\lbl{4.13}
\end{eqnarray}
satisfies
\begin{eqnarray}
      \gamma_{\mu} \cdot \gamma_{\nu} &=& \delta_{\mu \nu} \nonumber \\
      \gamma_{\bar \mu} \cdot \gamma_{\bar \nu} &=& - \delta_{\mu \nu}
      \nonumber \\
      \gamma_{\mu} \cdot \gamma_{\bar \mu} &=& 0
\lbl{4.14}
\end{eqnarray}
The numbers $(\gamma_{\mu},\, \gamma_{\bar \mu})$ thus form a set of
basis vectors of a vector space $V_{4,4}$ with signature $(+ + + + - - - -)$.

{\it 10-space.} \hs{3mm} Let $e_A = (e_{\mu}, e_{\bar \mu})$, 
$\mu = 1,2,3,4,5$; ${\bar
\mu} = {\bar 1}, {\bar 2}, {\bar 3}, {\bar 4},{\bar 5}$ be basis vectors of a 
10-dimensional Euclidean space $V_{10}$ with signature (+ + + ....).
We introduce ${\bar I} = e_{\bar 1} e_{\bar 2} e_{\bar 3} e_{\bar 4} 
e_{\bar 5}$ which satisfies
\begin{eqnarray}
         {\bar I}^2 &=& 1  \nonumber \\
         e_{\mu} {\bar I} &=& - {\bar I} e_{\mu} \nonumber \\
         e_{\bar \mu} {\bar I} &=& {\bar I} e_{\bar \mu}
\lbl{4.15}
\end{eqnarray}
Then the Clifford numbers
\begin{eqnarray}
       \gamma_{\mu} &=& e_{\mu} {\bar I}    \nonumber \\
       \gamma_{\bar \mu} &=& e_{\mu}
\lbl{4.16}
\end{eqnarray}
satisfy
     $$ \gamma_{\mu} \cdot \gamma_{\nu} = - \delta_{\mu \nu} $$
      $$\gamma_{\bar \mu} \cdot \gamma_{\bar \nu} =
       \delta_{{\bar \mu} {\bar \nu}} $$
\be       
       \gamma_{\mu} \cdot \gamma_{\bar \mu} = 0
\lbl{4.17}
\ee
The set $\gamma_A = (\gamma_{\mu}, \gamma_{\bar \mu})$ therefore spans
the vector space of signature $(- - - - - + + + + +)$.

The examples above demonstrate how vector spaces of various signatures are
obtained within a given set of polyvectors. Namely, vector spaces of
different signature are different subsets of polyvectors within the same
Clifford algebra.

This has important physical implications. We have argued that physical
quantities are polyvectors (Clifford numbers or Clifford aggregates). Physical
space is then not simply a vector space (e.g. Minkowski space), but a 
space of polyvectors. The latter is a pandimensional continuum
${\cal P}$ ($^{10}$) of points, 
lines, planes, volumes, etc., altogether. Minkowski
space is then just a sub-space with pseudo-Euclidean signature. Other sub-spaces
with other signatures also exist within the pandimensional continuum
${\cal P}$ and they all have physical significance. If we describe a particle
as moving in Minkowski spacetime $V_{1,3}$ we consider only certain
physical aspects of the considered object. We have omitted its other
physical properties like spin, charge, magnetic moment, etc. We can as well
describe the same object as moving in an Euclidean space $V_4$. Again such
a description would reflect only a part of the underlying physical
situation described by Clifford algebra.

\section{The unconstrained action\\
 from the polyvector action}

Let us consider the polyvector action (\ref{3.42}) and the constraint 
(\ref{2.6}). It is a polyvector equation, i.e. the sum of multivector
parts of different degrees. Each multivector part has to vanish separately.
Denoting the $r$-vector part as $\langle P^2 \rangle_r$, eq.(\ref{2.6}) can
be rewritten as a set of equations
\be
     \langle P^2 \rangle_r = 0 \; \; , \quad r = 0,1,2,3,4
\lbl{a1}
\ee
After some straightforward algebra we find
\be 
    \pi^{\mu} = 0 \; , \quad \mu = 0 \; , \quad S^{\mu \nu} = 0
\lbl{a2}
\ee
\be
      p^{\mu} p_{\mu} - m^2 = 0
\lbl{a3}
\ee
Therefore the polymomentum and the polyvelocity acquire the simplified
forms
\be
         P = p^{\mu} e_{\mu} + m e_5
\lbl{a4}
\ee
\be
        {\dot X} = {\dot x}^{\mu} + {\dot s} e_5
\lbl{a5}
\ee
and the action (\ref{3.42}) simplifies to the following phase space action
\be
     I[s,m,x^{\mu}, p_{\mu}, \lambda] =
     \int {\dd} \tau \left [ - m {\dot s} + p_{\mu} {\dot x}^{\mu} -
     {\lambda \oo 2} (p^{\mu} p_{\mu} - m^2) \right ]
\lbl{3.84}
\ee
which, besides $(x^{\mu},p_{\mu})$, has the additional variables
$(s,m)$.       

The equations of motion resulting from (\ref{3.84}) are
\begin{eqnarray}
    \delta s \, &:& \qquad {\dot m} = 0    \lbl{3.85} \\
    \delta m \, &:& \qquad {\dot s} - \lambda m = 0   \lbl{3.86} \\
    \delta x^{\mu} \, &:& \qquad {\dot p}_{\mu} = 0 \lbl{3.87} \\
    \delta p_{\mu}  \, &:& \qquad {\dot x}^{\mu} - \lambda p^{\mu} = 0
       \lbl{3.88} \\
    \delta \lambda  \, &:& \qquad  p^{\mu} p_{\mu} - m^2 = 0 \lbl{3.89}
\end{eqnarray}
We see that in this dynamical system the mass $m$ is one of the dynamical
variables; it is canonically conjugate to the variable $s$. From the
equations of motion we easily read out that $s$ is the proper time. Namely,
from (\ref{3.86}),(\ref{3.88}) and (\ref{3.89}) we have
\be
     p^{\mu} = {{{\dot x}^{\mu}} \oo \lambda} = m \, {{{\dd} x^{\mu}}\oo {{\dd}
     s}}
\lbl{3.90}
\ee
\be
    {\dot s}^2 = \lambda^2 m^2 = {\dot x}^2 \; \; , {\rm i.e} \quad
    {\dd} s^2 = {\dd} x^{\mu} {\dd} x_{\mu}
\lbl{3.91}
\ee
Using eq.(\ref{3.86}) we find that
\be
    - m {\dot s} + {\lambda \oo 2} m^2 = - {{m {\dot s}}\oo 2} = 
    - \, {1\oo 2} {{{\dd}(m s)}\oo {{\dd} \tau}} 
\lbl{3.92}
\ee
The action (\ref{3.84}) then becomes
\be
    I = \int {\dd} \tau \left ({1\oo 2} {{{\dd}(m s)}\oo {{\dd} \tau}} +
    p_{\mu} {\dot x}^{\mu} - {\lambda \oo 2} \, p^{\mu} p_{\mu} \right )
\lbl{3.93}
\ee
where $\lambda$ should be no longer considered as a quantity to be varied, 
but it is
now fixed: $\lambda = \Lambda(\tau)$. The total derivative in (\ref{3.93})
can be omitted, and the action is simply
\be
    I[x^{\mu}, p_{\mu}] = \int \dd \tau (p_{\mu} {\dot x}^{\mu} - 
    {\Lambda \oo 2}\, p^{\mu} p_{\mu})
\lbl{3.93a}
\ee
For a $\Lambda$  which is independent of $\tau$, (\ref{3.93a}) 
is just {\it the Stueckelberg action} ($^{2}$).
The equations of motion derived from (\ref{3.93a}) are
\be
     {\dot x}^{\mu} - \Lambda p^{\mu} = 0
\lbl{3.94}
\ee
\be
      {\dot p}_{\mu} = 0
\lbl{3.95}
\ee
From (\ref{3.95}) it follows that $p_{\mu} p^{\mu}$ is a constant of
motion. Denoting the latter constant of motion as $m$ and using (\ref{3.94})
we obtain that momentum can be written as
\be
    p^{\mu} = m \, {{{\dot x}^{\mu}}\oo {\sqrt{{\dot x}^{\nu}{\dot x}_{\nu}}}}
    = m \, {{{\dd} x^{\mu}}\oo {{\dd} s}} \; , \quad {\dd} s = ({\dd} x^{\mu}
    {\dd} x_{\mu})^{1/2}
\lbl{3.96}
\ee
which is the same as in eq.(\ref{3.90}). The equations of motion for
$x^{\mu}$ and $p_{\mu}$ derived from the Stueckelberg action (\ref{3.93a})
are the same as the equations of motion derived from the action (\ref{3.84}).
{\it A generic Clifford algebra action} (\ref{3.42}) {\it thus leads directly
to the Stueckelberg action}.

The above analysis can be easily repeated for a more general case,
by introducing a scalar constant $\kappa^2$, so that
instead of (\ref{3.84}) we have
\be
     I[s,m,x^{\mu}, p_{\mu}, \lambda] =
     \int {\dd} \tau \left [ - m {\dot s} + p_{\mu} {\dot x}^{\mu} -
     {\lambda \oo 2} (p^{\mu} p_{\mu} - m^2 - \kappa^2) \right ]
\lbl{3.97}
\ee
Then, instead of (\ref{3.93a}), we obtain
\be
     I[x^{\mu}, p_{\mu}] = \int {\dd} \tau \left ( p_{\mu} {\dot x}^{\mu}
     - {\Lambda \oo 2} \, (p^{\mu} p_{\mu} - \kappa^2) \right )
\lbl{3.98}
\ee
The corresponding Hamiltonian is
\be
    H = {\Lambda \oo 2} \, (p^{\mu} p_{\mu} - \kappa^2) 
\lbl{3.98a}
\ee
and in the quantized theory the Schr\" odinger equation reads
\be
     i {{\p \psi} \oo {\p \tau}} =  
     {\Lambda \oo 2} \, (p^{\mu} p_{\mu} - \kappa^2) \psi
\lbl{3.98b}
\ee

If we derive from (\ref{3.97}) the equations of motion (which are
straightforward generalizations, for $\kappa \neq 0$, of 
eqs.(\ref{3.85})-(\ref{3.89})), and eliminate the conjugate variables
$p_{\mu}$ and $m$, we can re-express the action (\ref{3.97}) as
\be
       I[x^{\mu},s] = \kappa \int \dd \tau ({\dot x}^{\mu} {\dot x}_{\mu}
       - {\dot s}^2)^{1/2}
\lbl{a6}
\ee
This is a straightforward generalization of the usual relativistic 
point particle action to an extra variable $s(\tau)$. It is important to
bear in mind that this extra variable $s$ is not due to a postulated
extra dimension of spacetime, but due to the existence of the Clifford
algebra generated by the basis vectors of spacetime. Although spacetime
remains 4-dimensional, a point particle is described not only by
four coordinates variables $x^{\mu}(\tau)$, but also by an extra variable
$s(\tau)$.

The extra variable $s$ has brought us to what appears (in the specific case
considered) as an $O(1,4)$ invariant action. 
The ``$O(1,4)$" action contains the constraint, therefore the
extra variable is not a variable at all (at least if we choose the remaining
ones -- i.e., $x^{\mu}$ -- as the true variables). The extra variable s
is related to the parameter $\tau$ through a choice of "gauge",
that is by choice of the Lagrange multiplier $\lambda$. In
the particular case we first chose $\lambda = \Lambda(\tau)$. Further
we have chosen $\Lambda(\tau)$ as independent of $\tau$. Then one
finds $s = \Lambda m \tau$. In such a choice of parametrization
$s$ is proportional to $\tau$. Other parametrizations are, of course,
possible, and in this respect the ``$O(1,4)$" action goes beyond
Stueckelberg. But physically it is equivalent to the Stueckelberg
action, because an arbitrary choice of gauge (parametrization)
has no influence on physics. In short, if we choose $x^{\mu}$
as the dynamical variables (evolution in spacetime, relativistic
dynamics), then the $s$ is not a variable at all\footnote{Analogously, in
the usual theory of relativity, because of the mass shell constraint,
$x^0 \equiv t$ is not a variable at all, and it is often considered as
the evolution parameter: so one obtains the evolution in 3-space,
but not in spacetime $V_{1,3}$.}; it can be chosen
to be equal, or at least proportional to $\tau$. 
And most important, the "$O(1,4)$" action does not come from a space $V_{1,4}$,
but from the Clifford algebra over $V_{1,3}$. 

\section{The polyvector action and DeWitt--Rovelli material reference fluid}

In a remarkable paper ($^{17}$) Rovelli considered in modern language the
famous Einstein ``hole argument" which shows that points of empty
spacetime cannot be identified. For a precise formulation the reader is adviced
to have a look at Rovelli's paper. Here I present the argument, as I
understand it, in a simplified and compact way.

We are familiar with the fact that the Einstein equations are invariant
under general coordinates transformations. In a given coordinate system
$O$, let $g_{\mu \nu} (x), \, X_i^{\mu} (\tau)$ be a solution to the
Einstein equations - a possible universe $U$, with the metric $g_{\mu \nu} (x)$
and the set of point particles' world lines $X_i^{\mu} (\tau)$, $i = 1,2,...,N$.
The same universe $U$ can be expressed in a different coordinate system
$O'$ as $g'_{\mu \nu} (x'), \, {X'}_i^{\mu} (\tau')$ which, of course, is
also a solution to the Einstein equations. This transformation is called
also a {\it passive diffeomorphism}.

Let us now consider another kind of transformation, namely an {\it active
diffeomorphism} which, in the same coordinate system $O$ sends a universe
$U$, described by $g_{\mu \nu} (x), \, X_i^{\mu} (\tau)$ into another
universe $U'$, described by $g'_{\mu \nu} (x), \, {X'}_i^{\mu} (\tau)$.
There is a lot of freedom in choosing active diffeomorphisms. Can then the
universes $U$ and $U'$ be physically distinct?

The same initial conditions should lead to the same physical universes. But
active diffeomorphisms allow for the possibility that, starting from 
the same initial conditions at a given spacelike hypersurface (where
$U$ and $U'$ are identical), we can arrive at the situation where $U$ and
$U'$ are distinct at a ``later" spacelike hypersurface. If $U$ and $U'$
were {\it physically} distinct, then determinism would be violated. Hence
$U$ and $U'$ must be physically the same (even if described by different
sets of variables related by an active diffeomorphism). But, being the same,
spacetime points in the {\it holes} within matter configuration (the latter
being described
by the set of worldlines $X_i^{\mu} (\tau)$) cannot be identified.

If we wish to build up a theory in which spacetime points could be identified,
we have to fill spacetime  with a {\it reference fluid}. Such an idea
was earlier considered by DeWitt ($^{18}$), and revived by Rovelli
($^{17}$). As a starting point Rovelli considers a simplified 
{\it reference system} consisting of a {\it single particle} and a {\it clock}
attached to it. Variables are then particle's coordinates $X^{\mu} (\tau)$
and {\it the clock variable} $T(\tau)$. As a model of general relativity
+ material reference system theory Rovelli considers the action whose matter
part is
\be
       I = m \int \dd \tau \left ( {{\dd X^{\mu}}\oo {\dd \tau}} \, 
       {{\dd X_{\mu}}\oo {\dd \tau}} - {1\oo \omega^2} \left ({{\dd T}\oo
       {\dd \tau}} \right )^2 \right )^{1/2}
\lbl{a7}
\ee

If we make replacement $m \rightarrow \kappa$, $T/\omega \rightarrow s$,
we obtain precisely the action (\ref{a6}) derived from the polyvector
action. Our polyvector action can be generalized ($^{13}$) to strings 
and higher dimensional membranes ($p$-branes). We obtain the {\it unconstrained
action} starting from the constrained action which includes the 
{\it pseudoscalar field}. The latter field is a necessary 
ingredient of the polyvector
generalization of the theory. On the other hand, DeWitt and Rovelli have taken
a fluid of reference particles and obtained a similar action which involves
a {\it field of the clock variable}.

\section{Conclusion}

Clifford algebra is an immensely useful language for geometry and physics. It
contains quaternions and differential forms as special cases. Equations of
physics acquire remarkably condensed forms. There is a lot of room for
new physics. It illuminates the role of spinors: they are a special kind
of polyvectors. Clifford algebra, together with the conception of
physical quantities as polyvectors (Clifford aggregates), is very likely
the language of a future unified theory. What I was able to present here
is only a tip of an iceberg\footnote{A slightly greater part of the iceberg
is uncovered in ref. ($^{13}$).}

Geometric calculus based on Clifford algebra leads to the point particle action
with an extra variable --the clock variable-- which enables
evolution in spacetime. In other words, our model with the polyvector action
allows for the {\it dynamics} in spacetime (relativistic dynamics). 
Relativistic
dynamics is a necessary consequence of the existence of Clifford algebra
as a general tool for description of geometry of spacetime. Moreover,
when considering {\it dynamics of spacetime itself}, such  model,
in my opinion, provides a natural resolution of ``the problem of time"
in quantum gravity. A number of researchers have come close to the
viewpoint that even in gravity one has to introduce an extra, invariant,
parameter which serves the role of evolution time
($^{19,20,21}$). The latter parameter, as
already stated, in the polyvector generalization of physics is not postulated
but is present automatically.

\vs{1cm}

\centerline{\bf Ackonwledgement}

\vs{4mm}

One of the turning points in my work was
when in 1992 I met prof. Waldyr Rodrigues, Jr. We were both guests
of Erasmo Recami at Istituto di Fisica Teorica, Catania, Italy. Our aim
was to collaborate in a joint project on various models of the spinning particle
and find the connection between the Barut-Zanghi ($^{22}$)
model and its reformulation
by means of Clifford algebra. So prof. Rodrigues started to talk me about
Clifford algebra as a useful tool for geometry and physics.
After two weeks of discussion I became
a real enthusiast of the Clifford algebra. In this paper I wished to 
forward my enthusiasm to those readers who are not yet enthusists
themselves.

The work was supported by the Slovenian Ministry of Science and Technology.

\newpage

\centerline{\bf \large References}

\bigskip

\begin{enumerate}

\item V. Fock, {\it Phys. Z. Sowj.} {\bf 12}, 404 (1937)

\item E.C.G. Stueckelberg, {\it Helv. Phys. Acta}, 
{\bf 14}, 322 (1941); {\bf 14}, 588 (1941); {\bf 15}, 23 (1942)

\item R. P. Feynman {\it Phys. Rev}, {\bf 84}, 108 (1951)

\item J. Schwinger, {\it Phys. Rev}, {\bf 82}, 664 (1951)

\item W. C. Davidon, {\it Physical Review} {\bf 97},1131 (1955);
{\bf 97},1139 (1955)

\item L. P. Horwitz and C. Piron, {\it Helv. Phys. Acta}, {\bf 46}, 316 (1973);
L. P. Horwitz and F. Rohrlich, {\it Physical Review D} {\bf 24}, 1528 (1981);
{\bf 26}, 3452 (1982); L. P. Horwitz, R. I. Arshansky and A. C. Elitzur
{\it Found. Phys} {\bf 18}, 1159 (1988); R. Arshansky, L. P. Horwitz and
Y. Lavie, {\it Foundations of Physics} {\bf 13}, 1167 (1983);
L. P. Horwitz, in {\it Old and New Questions in Physics, Cosmology,
Philosophy and Theoretical Biology} (Editor Alwyn van der Merwe, Plenum,
New York, 1983); L. P. Horwitz and Y. Lavie, {\it Physical Review D} {\bf 26},
819 (1982);
L. Burakovsky, L. P. Horwitz and W. C. Schieve, {\it Physical Review D}
{\bf 54}, 4029 (1996); L. P. Horwitz and W. C. Schieve,
{\it Annals of Physics} {\bf 137}, 306 (1981)

\item J.R.Fanchi, {\it Phys. Rev. D} {\bf 20}, 3108 (1979);
see also the review J.R.Fanchi, {\it Found. Phys.} {\bf 23},
287 (1993), and many references therein; J. R. Fanchi {\it Parametrized
Relativistic Quantum Theory} (Kluwer, Dordrecht, 1993)

\item H.Enatsu, {\it Progr. Theor. Phys} {\bf 30}, 236 (1963); 
{\it Nuovo Cimento A} {\bf 95}, 269 (1986);
F. Reuse, {\it Foundations of Physics} {\bf 9}, 865 (1979);
A. Kyprianidis {\it Physics Reports} {\bf 155}, 1 (1987);
R. Kubo, {\it Nuovo Cimento A} {\bf}, 293 (1985);
M. B. Mensky and H. von Borzeszkowski, {\it Physics Letters A}
{\bf 208}, 269 (1995); J. P. Aparicio, F. H. Gaioli and E. T. Garcia-Alvarez,
{\it Physical Review A} {\bf 51}, 96 (1995); {\it Physics Letters A}
{\bf 200}, 233 (1995); L. Hannibal, {\it International Journal of
Theoretical Physics} {\bf 30}, 1445 (1991);
F. H. Gaioli and E. T. Garcia-Alvarez, {\it General Relativity and
Gravitation} {\bf 26},1267 (1994) 

\item M. Pav\v si\v c, {\it Found. Phys.} {\bf 21}, 1005 (1991);
M. Pav\v si\v c,{\it Nuovo Cim.} {\bf A104}, 1337 (1991); 
{\it Doga, Turkish Journ. Phys.} {\bf 17}, 768 (1993)

\item W. M. Pezzaglia Jr, {\it Classification of 
Multivector Theories and
Modification o f the Postulates of Physics}, e-Print Archive: gr-qc/9306006;\\ 
W. M. Pezzaglia Jr, {\it Polydimensional Relativity, a Classical 
Generalization of the
Automorphism Invariance Principle}, e-Print Archive: gr-qc/9608052;\\
W. M. Pezzaglia Jr, {\it Physical Applications of a Generalized Clifford
Calculus: Papapetrou Equations and Metamorphic Curvature},
e-Print Archive: gr-qc/9710027;\\  
W. M. Pezzaglia Jr nad J. J. Adams, {\it Should Metric Signature Matter 
in Clifford Algebra Formulation of
Physical Theories?}, e-Print Archive: gr-qc/9704048;\\   
W. M. Pezzaglia Jr and A. W. Differ, {\it A Clifford Dyadic Superfield from
Bilateral Interactions of Geometric Multispin Dirac Theory},
e-Print Archive: gr-qc/9311015;\\
W. M. Pezzaglia Jr, {\it Dimensionally Democratic Calculus and
Principles of Polydimensional Physics}, e-Print Archive: gr-qc/9912025

\item C. Castro, {\it The String Uncertainty Relations follow from
the New Relativity Principle}, e-print Archive: hep-th/0001023;\\
C. Castro, {\it Is Quantum Spacetime Infinite Dimensional?},
e-Print Arhive: hep-th/0001134;\\
C. Castro, {\it Chaos, Solitons and Fractals} {\bf 11}, 1721 (2000);\\
C. Castro and A. Granik,  {\it On M Theory, Quantum Paradoxes and the New
Relativity}, e-print Archive: physics/0002019;

\item M. Pav\v si\v c, ``Clifford Algebra as a Useful Language
for Geometry and Physics", in {\it Geometry and Physics}, Proccedings of the
38. Internationale Universit\" atswochen f\" ur Kern- und Teilchenphysik,
Schladming, Austria, January 9-16,1999 (Editors
H. Gauster, H. Grosse and L. Pittner, Springer, Berlin, 2000)

\item M. Pav\v si\v c, {\it The Landscape of Theoretical Physics :
A Global View} (Kluwer Academic, to appear)

\item
D. Hestenes, {\it Space-Time Algebra} (Gordon and Breach,
New York, 1966); D. Hestenes {\it Clifford Algebra to Geometric Calculus}
(D. Reidel, Dordrecht, 1984)

\item S. Teitler, {\it Supplemento al Nuovo Cimento} {\bf III}, 1 (1965);
{\it Supplemento al Nuovo Cimento} {\bf III}, 15 (1965);
{\it Journal of Mathematical Physics} {\bf 7}, 1730 (1966);
{\it Journal of Mathematical Physics} {\bf 7}, 1739 (1966)

\item L. P. Horwitz, {\it J. Math. Phys.} {\bf 20}, 269 (1979);
H. H. Goldstine and L. P. Horwitz, {Mathematische Annalen} {\bf 164},
291 (1966)

\item C. Rovelli, {\it Classical and Quantum Gravity} 
{\bf 8}, 297 (1991); {\bf 8} 317 (1991)

\item B. S. DeWitt, in {\it Gravitation: An Introduction to Current
Research} (Editor L. Witten, Wiley, New York, 1962)

\item M. Pav\v si\v c, {\it Foundations of  Physics} {\bf 26}, 159 (1996)

\item J. Greenstie, {\it Classical and Quantum Gravity} {\bf 13},
1339 (1996); {\it Physical Review D} {\bf 49}, 930 (1994);
A. Carlini and J. Greensite, {\it Physical Review D} {\bf 52},
936 (1995); {\bf 52}, 6947 (1955);  {\bf 55}, 3514 (1997);

\item J. Brian and W. C. Schieve, {\it Foundations of Physics} {\bf 28},
1417 (1998)

\item A. O. Barut and N. Zanghi, {\it Phys. Rev. Lett.} {\bf 52},
2009 (1984)

\end{enumerate}

\end{document}